\documentclass[12pt]{article}
 
\usepackage{epsfig}
 
\newcount\xrefpos \xrefpos=0
\newcount\yrefpos \yrefpos=0
\newcount\xput \xput=0
\newcount\yput \yput=0
\def\refpos#1 #2 #3{\global\xrefpos=#1 \global\yrefpos=#2
                         \rlap{$\smash{#3}$}}
\def\put #1 #2 #3{\xput=#1 \yput=#2
                  \advance\xput by -\xrefpos
                  \advance\yput by -\yrefpos
                  \rlap{\kern\the\xput truebp
                        \vbox to 0pt{\vss\hbox{$\displaystyle #3$}
                        \kern\the\yput truebp}}}
\def\beginlabels\refpos#1\endlabels{\hbox{$\refpos#1$}}
 
\usepackage{amsfonts,amssymb}
 
\newcommand{\sect}[1]{\setcounter{equation}{0}\section{#1}}

\newcommand{\al}{\ensuremath{\alpha}}

\newcommand{\Ga}{\ensuremath{\Gamma}}

\newcommand{\ep}{\ensuremath{\epsilon}}

\newcommand{\la}{\ensuremath{\lambda}}

\newcommand{\om}{\ensuremath{\omega}}
\newcommand{\Om}{\ensuremath{\Omega}}
\newcommand{\p}{\ensuremath{\phi}}

\renewcommand{\th}{\ensuremath{\theta}}

\renewcommand{\L}{\ensuremath{{\cal L}}}

\renewcommand{\d}{\ensuremath{{\rm d}}}

\newcommand{\del}{\ensuremath{\partial}}

\newcommand{\td} {\ensuremath{\tilde}}

\newcommand{\be}{\begin{equation}}
\newcommand{\ee}{\end{equation}}
 
\newcommand{\ba}{\begin{eqnarray}}
\newcommand{\ea}{\end{eqnarray}}
 
\begin{document}
 
\bigskip
\hskip 4.8in\vbox{\baselineskip12pt \hbox{hep-th/0306190}}
 
\bigskip
\bigskip
\bigskip
 
\begin{center}
{\Large \bf Closed Timelike Curves and Holography}\\
\bigskip
{\Large \bf in Compact Plane Waves}
\end{center}
 
\bigskip
\bigskip
\bigskip
 
\centerline{\bf D. Brecher$^\natural$, P. A. DeBoer$^\natural$, D.
C. Page$^\flat$ and M. Rozali$^\natural$}
 
\bigskip
\bigskip
\bigskip
 
\centerline{\it $^\natural$Department of Physics and Astronomy}
\centerline{\it University of British Columbia} \centerline{\it
Vancouver, British Columbia V6T 1Z1, Canada} \centerline{\small
\tt brecher, pdeboer, rozali@physics.ubc.ca}
 
\centerline{$\phantom{and}$}
 
\centerline{\it $^\flat$Department of Physics and Astronomy}
\centerline{\it University of Toronto, Toronto, Ontario M5S 1A7,
Canada} \centerline{\small \tt page@physics.utoronto.ca}
 
\bigskip
\bigskip
 
\begin{abstract}
\vskip 2pt We discuss plane wave backgrounds of string theory and
their relation to G\"{o}del--like universes.  This involves a
twisted compactification along the direction of propagation of the
wave, which induces closed timelike curves.  We show, however,
that no such curves are geodesic.  The particle geodesics and the
preferred holographic screens we find are qualitatively different
from those in the G\"{o}del--like universes.  Of the two types of
preferred screen, only one is suited to dimensional reduction
and/or T--duality, and this provides a ``holographic
protection'' of chronology.  The other type of screen, relevant to an
observer localized in all directions, is constructed both for the compact and
non--compact plane waves, a result of possible independent interest.
We comment on the consistency of field theory in such spaces, in which
there are closed timelike (and null) curves but no closed timelike (or
null) geodesics.
\end{abstract}
 
\newpage
 
\baselineskip=18pt \setcounter{footnote}{0}


\sect{Introduction}
 
The four--dimensional G\"{o}del universe~\cite{godel:49} is a
topologically trivial homogeneous space with non--zero rotation.
This gives rise to some unusual properties: not only do all
observers see themselves as the centre of rotation, but there
exist closed timelike curves (CTCs) through every point.  As an
example of a space with CTCs for all times, it is unclear as to
what extent Hawking's chronology protection
conjecture~\cite{hawking:92}, concerning the impossibility of
\emph{forming} a CTC in nature, is applicable.
 
For this reason, the discussion of the physics of chronology
protection (see, \emph{e.g.},~\cite{visser:02} for a recent
review), has mostly avoided the G\"{o}del universe.  However, the
discovery~\cite{gauntlett:02} of supersymmetric G\"{o}del--like
solutions to five--dimensional minimal supergravity, and to its
eleven--dimensional M--theoretic lift, has forced the issue, at
least within the string theory community.  Surprisingly, these
solutions to string and M--theory turn
out~\cite{boyda:02,harmark:03} to be related to another
supersymmetric space of much recent interest, namely the plane
wave~\cite{blau:01,metsaev:01,berenstein:02}, and this is the
relation of interest to us here.  We should also note that
supergravity solutions describing supersymmetric deformations of
the extreme five--dimensional Reissner--Nordstrom black hole have
been studied in this context.  Both of the deformations discussed
in~\cite{herdeiro:02} --- one corresponding to the rotating black
hole of~\cite{breckenridge:96}, the other to a black hole in a
G\"{o}del--like universe --- have CTCs, the precise nature of
which has been examined in some
detail~\cite{gibbons:99,herdeiro:00,herdeiro:02,jarv:02,dyson:03}.
Other non--supersymmetric solutions of interest, describing black
holes in a G\"{o}del--like universe, have recently been discussed
in~\cite{gimon:03}.
 
It turns out that these G\"{o}del--like universes (GLUs) are
related to \emph{compactified} plane waves (CPWs) in two different
ways.  The eleven--dimensional CPW, dimensionally reduced on an
everywhere spacelike circle, yields a ten--dimensional
GLU~\cite{harmark:03}. On the other hand, a CPW in type II string
theory is T--dual to a GLU times a transverse
circle~\cite{boyda:02}.  Depending on the specific plane wave we
start with, these procedures give rise to a large variety  of
GLUs.  In addition to the R--R fluxes typically supporting the CPW
solutions, dimensionally reducing a CPW gives rise to a magnetic
R--R one--form potential, whereas T--dualizing gives instead a
NS--NS two--form potential.  These background fluxes are needed to
preserve supersymmetry, although the specific number of
supersymmetries preserved is not invariant under the dimensional
reduction or T--dualization.
 
In this paper we concentrate on the CPW side of the dual pair. We
show in section 2 that there are CTCs in this set of geometries
though, as for the GLU~\cite{boyda:02}, these are not geodesic.
This is a fact we find significant with respect to the consistency
of propagation in this set of backgrounds.  We proceed to
construct, in section 3, the geodesics in the CPW background.
These are needed in order to identify the holographic
screens~\cite{bousso:9905,bousso:9906} in the CPW geometry.
 
In section 4 we discuss the construction of lightsheets and of
preferred holographic screens, in both compact and non--compact plane wave
geometries.  In both cases the lightsheets constructed are
bounded by spatial surfaces, and the covariant entropy
bound~\cite{bousso:9905,bousso:9906} can
be applied.  Constructing a preferred holographic screen entails taking
the union of these surfaces.  In the non--compact case, this is a
sensible thing to do and we exhibit the result, which may be of some
interest in the study
of plane wave holography~\cite{pp1}--\cite{pp6}.  In the compact
case, however, we find that the region
``enclosed'' within the holographic screen is the complete spacetime.
This is related to the existence of a compact direction, and not
necessarily to the existence of CTCs.  The interesting mechanism of
``holographic protection'' of chronology does not appear to be operating in this
case --- although
all CTCs intersect the screen, they are also ``enclosed'' within it,
in the precise sense defined by Bousso~\cite{bousso:9905,bousso:9906}.

In section 5, we comment on the relation to the similar analyses
of the GLU performed in~\cite{boyda:02}.  There are, in
fact, two types of preferred holographic screen in our geometry.
In addition to the ones discussed above, there are screens
associated with the Kaluza--Klein zero modes which are ``smeared''
along the compact direction, just as in the GLU with flat
transverse directions~\cite{boyda:02}.  This smeared screen is the
origin of the holographic screen in the GLU and we argue that
similar statements are true in any backgrounds related by
Kaluza--Klein dimensional reduction and/or T--duality.
 
Of course, in non--perturbative string theory, the GLUs and the
CPWs are \emph{two descriptions of a single object}.  However, the
approach to holography reviewed in~\cite{bousso:02} separates out
the string fields into ``metric'' and ``matter fields'', as in
traditional semiclassical approaches to quantum gravity.    The
interplay between the backgrounds described here provides
hints as to how to extend the notion of holography to string
backgrounds with fluxes inherited from a higher--dimensional or
T--dual metric.  We discuss  how the covariant entropy bounds
of~\cite{bousso:9905,bousso:9906} can be  affected by dimensional
reduction and/or T--duality\,\footnote{For a possible relation of
holography with  supersymmetry, see~\cite{tom1,tom2}.}.
 
We conclude in section 6 with comments on attempts to quantize
particles and fields in the CPW, and the related issues in the
GLU.  Since the only closed geodesics in these geometries are
spacelike, some of the obvious problems with defining field theory
in spaces with closed timelike and/or null geodesics are avoided.
In this respect, quantum field theory might be well behaved in
those backgrounds. For a recent discussion of potential problems
with the  GLU in string theory, see~\cite{joan}.

While preparing
this manuscript for publication, the preprint~\cite{hikida:03}
appeared, which includes some results similar to those of section 3.


\sect{Closed Timelike Curves}
 
In this section we introduce the basic spacetimes --- a class of
plane waves with a compact direction (CPWs), and the related
G\"{o}del--like universes (GLUs). We show that both these
geometries have CTCs (as well as closed null curves). Since they
are homogeneous spaces, both have CTCs through every point. We
also show that none of the CTCs is a geodesic, thus avoiding some
of the obvious problems such curves would generate.
 
\subsection{Compactified Plane Waves}
 
The class of metrics we are interested in can be written as \be \d
s^2_D = \d u \d v - \sum_{i=1}^N \beta_i^2 \rho_i^2 \d u^2 +
\sum_{i=1}^N \left( \d \rho_i^2 + \rho_i^2 \d \td{\p}_i^2 \right)
+ \sum_{\al=1}^{D-2(N+1)} \d x^\al \d x^\al, \label{eqn:metric1}
\ee
 
\noindent where $\{\rho_i,\td{\p}_i\}$ are polar coordinates in
$N$ transverse planes, and the light--cone coordinates are $u = z
+ t, \,\, v = z - t.$  We have included as many flat directions as
necessary to make this a background of string ($D=10$) or
M--theory ($D=11$).
 
Note that the metric is not the pure gravitational plane wave, and
must be supported by various form fields.  Furthermore, for
$D=10$, this class of backgrounds gives rise to (mass)$^2$ terms
on the string worldsheet which are all positive. String theory on
such backgrounds is well understood~\cite{metsaev:01,metsaev:02},
as are the effects of compactification of various spacelike
circles~\cite{michelson:02}.
 
The relationship between the plane wave and the associated GLU
involves dimensional reduction or T--duality along a spatial
coordinate~\cite{boyda:02,harmark:03}.  This coordinate cannot be
simply $z$,  since the Killing vector \be \xi_0 = \frac{\del}{\del
z} = \frac{\del}{\del u} + \frac{\del}{\del v}, \ee
 
\noindent is not everywhere spacelike: \be \left|\xi_0\right|^2 =
1 - \sum_{i=1}^N \beta_i^2 \rho_i^2. \ee
 
\noindent Instead we can use~\cite{boyda:02} the Killing vector
\be \xi = \xi_0 - \sum_{i=1}^N \beta_i \frac{\del}{\del \td{\p}_i}
\qquad \Rightarrow \qquad \left| \xi \right|^2 =  1, \ee
 
\noindent which \emph{is} everywhere spacelike\,\footnote{Such
twisted compactifications of solutions of M--theory have been much
studied~\cite{farrill:01,farrill:0208107,farrill:0208108},
although this classification has precisely precluded reductions
which give rise to CTCs.}.
 
Identifying points a distance $2\pi nR$ along the orbits of $\xi$
involves the following identifications on the coordinates: \be (z,
\td{\p}_i) \equiv (z + 2\pi n R, \td{\p}_i + 2\pi m_i - 2\pi n
R\beta_i). \label{eqn:identify} \ee
 
\noindent Since $\xi$ is everywhere spacelike we can compactify or
T--dualize along its orbits.  To facilitate this, we define the
adapted coordinates \be \p_i = \td{\p}_i + \beta_i u,
\label{eqn:rotation} \ee
 
\noindent which are constant along orbits of $\xi$: $\xi(\p) = 0$.
The $z$--rotation ensures that the $\p_i$ have standard
periodicity, the identifications becoming \be (z, \p_i) \equiv (z
+ 2\pi nR, \p_i + 2 \pi m_i). \ee
 
\noindent In these coordinates, the metric
is~\cite{boyda:02,harmark:03} \be \d s^2_D = \d u \d v - 2
\sum_{i=1}^N \beta_i \rho_i^2 \d \p_i \d u + \sum_{i=1}^N \left(
\d \rho_i^2 + \rho_i^2 \d \p_i^2 \right) + \sum_{\al=1}^{D-2(N+1)}
\d x^\al \d x^\al, \label{eqn:metric2} \ee
 
\noindent where $\p_i$ are angles with standard periodicity, and
the Killing vector $\xi$ is simply \be \xi = \frac{\del}{\del z} =
\frac{\del}{\del u} + \frac{\del}{\del v}. \ee
 
\noindent Dimensional reduction or T--duality along $z$ is now
straightforward, and will be reviewed below.
 
We now show that there are CTCs (and closed null curves) in the
above CPW, and that  none of them are geodesic.  Consider the
curve generated by the Killing vector \be K = \frac{\del}{\del z}
+ \sum_{i=1}^N \al_i \frac{\del}{\del \p_i}, \ee where $\al_i$ are
constant.
 
This curve is closed if and only if the $\alpha_i$ are rational
multiples of $1/R$ for each $i$\,\footnote{To see this, put the
$N$ rational multiples over a common denominator $q$: $ \al_i =
p_i/(q R), \,\, p_i, q \in \mathbb{Z}.$  Then if we travel from
$z$ to $z + 2 \pi q R$ along the curve we get back to where we
started: $ (z, \p_i) \equiv (z + 2 \pi q R, \p_i + 2 \pi p_i)$.}.
Furthermore, this curve is timelike for sufficiently large
$\rho_i$ if $\alpha_i < 2 \beta_i$.  This follows
straightforwardly from the metric (\ref{eqn:metric2}).  The CTCs
exist for \be \rho_i^2 > \frac{1}{(2 \beta_i - \alpha_i) \alpha_i}
\ge \frac{1}{ \beta_i^2}, \label{eqn:ctc} \ee
 
\noindent the latter inequality saturated for $\alpha_i =
\beta_i$.  By picking a rational value of $\alpha_i$ arbitrarily
close to $\beta_i$ we find CTCs arbitrarily close to $\rho_i =
\beta_i^{-1}$ for any value of $\beta_i$.  We therefore conclude
that the CPWs considered here develop CTCs beyond the critical
radii\,\footnote{The surfaces of $\rho=\rho_{\rm critical}$ are
referred to as ``velocity of light surfaces'' in~\cite{boyda:02}.}
$\rho_i = \beta_i^{-1}$.  Note that closed null curves (CNCs)
develop at the critical radii themselves, provided $\beta_i$ is a
rational multiple of $1/R$.
 
It is easy to see that none of these CNCs or CTCs are geodesic
since, with a dot denoting differentiation with respect to an
affine parameter, \be \dot{K}^{\rho_i} + \Ga^{\rho_i}_{~bc} K^b
K^c =  \rho_i \al_i (2\beta_i - \al_i), \ee all other components
of the geodesic equation being trivial.  Only those curves with
$\al_i=0$ or $\al_i=2\beta_i$ (for all values of $i$) are geodesic
and in these cases, we see from (\ref{eqn:ctc}) that the CTCs are
pushed off to infinity. One can, in fact, prove more generally
that there are no closed timelike or null geodesics in the CPW,
but we will postpone this proof until section \ref{sec:geodesics},
since we will need the explicit solution to the geodesic
equations.
 
\subsection{The  G\"{o}del--Like Universe}
 
The CPW is related~\cite{boyda:02,harmark:03} to a GLU by the
closely related operations of T--duality or dimensional reduction
along the orbits of the Killing vector $\xi$ above.  This is most
easily achieved by rewriting the metric (\ref{eqn:metric2}) in a
form adapted to dimensional reduction along the orbits of
$\del/\del z$: \be \d s^2_D = - \left( \d t + \om \right)^2 +
\sum_{i=1}^N \left( \d \rho_i^2 + \rho_i^2 \d \p_i^2 \right) +
\sum_{\al=1}^{D-2(N+1)} \d x^\al \d x^\al + \left( \d z - \om
\right)^2, \label{eqn:reduce} \ee
 
\noindent where we  define the one--form \be \om = \sum_i \beta_i
\rho_i^2 \d \p_i. \ee
 
\noindent We therefore identify the dimensionally reduced metric
as~\cite{harmark:03} \be \d s^2_{D-1} = - \left( \d t + \om
\right)^2 + \sum_{i=1}^N \left( \d \rho_i^2 + \rho_i^2 \d \p_i^2
\right) + \sum_{\al=1}^{D-2(N+1)} \d x^\al \d x^\al, \ee
 
\noindent describing a $(2N+1)$--dimensional GLU times
$\mathbb{E}^{D-2(N+1)}$, together with an R--R 1-form potential
given by $C=-\om$.
 
If we think instead of the metric (\ref{eqn:metric2}) as a CPW of
type II string theory, one can T--dualize along orbits of
$\del/\del z$, giving~\cite{boyda:02,harmark:03} \be \d s^2_D = -
\left( \d t + \om \right)^2 + \sum_{i=1}^N \left( \d \rho_i^2 +
\rho_i^2 \d \p_i^2 \right) + \sum_{\al=1}^{D-2(N+1)} \d x^\al \d
x^\al + \d z^2, \ee
 
\noindent which is just the same GLU space as above, times the
T--dual circle. In addition  there is a NS--NS two--form potential
$B = \d z \wedge \om$.  The theory reduced on this T--dual circle
is identical to the one obtained by direct reduction of the plane
wave metric, since T--duality is a symmetry of the dimensionally
reduced theory.  The solutions which one can generate in these
ways have been extensively classified in~\cite{harmark:03}.
 
Either way, the resulting metric clearly has CTCs generated by
$\del/\del \p_i$ and, in both cases, some background fields are
generated.  These are required for unbroken supersymmetry.
 
We will see that  the background fields change the behaviour of
charged particles and fields.  The charged objects are simply
momentum modes in the CPW, and they become  D0-branes if one
dimensionally reduces an M--theory solution, or string winding
modes in the T--dual GLU$\times S^1$ geometry of type II string
theory.


\sect{Geodesics} \label{sec:geodesics}
 
Next we discuss the geodesics in the CPW, following closely the
analysis of~\cite{boyda:02}.  These include the particle geodesics
discussed in the T--dual picture in~\cite{boyda:02}, but also
trajectories of charged particles --- which would be string
winding modes in the discussion of~\cite{boyda:02}.  We will see
that the latter behave in a way qualitatively different to that of
the uncharged geodesics.
 
The Lagrangian for a particle moving in the metric
(\ref{eqn:metric2}) (with a dot denoting differentiation with
respect to the affine parameter $\la$) is \be \L = \dot{u} \dot{v}
- 2 \sum_i \beta_i \rho_i^2 \dot{\p}_i \dot{u} + \sum_i \left(
\dot{\rho}_i^2 + \rho_i^2 \dot{\p}_i^2 \right)+P^2  = -\ep, \ee
 
\noindent where $\ep = 0$ for a null geodesic, $\ep = 1$ for a
timelike geodesic and $\ep = -1$ for a spacelike geodesic.  We
have also substituted for the $(D-2(N+1))$ conserved momenta,
$P_\al$, in the flat directions, with $P^2 = \sum_{\al} P_\al^2$.
 
There are a further $N+2$ conserved quantities coming from the
Killing vectors $\del/\del u$, $\del/\del v$ and $\del/\del \p_i$,
the associated geodesic equations being \be \frac{\del \L}{\del
\dot{v}} = P_- \equiv \frac{1}{2} (P_z + E), \qquad \frac{\del
\L}{\del \dot{u}} = P_+ \equiv \frac{1}{2} (P_z - E), \qquad
\frac{\del \L}{\del \dot{\p}_i} = 2 L_i, \ee
 
\noindent where, since $z$ is compact, we will ultimately be
interested in taking $P_z=n$ for some integer $n$.  Inverting
these relations gives \be \dot{u} = P_-, \qquad \dot{v} = P_+ +
2\sum_i  \left( \beta_i L_i  + P_-\beta_i^2 \rho_i^2 \right),
\qquad \dot{\p}_i = \frac{L_i}{ \rho_i^2} + \beta_i P_-.
\label{relations} \ee
 
\noindent and the resulting constraint for the radial coordinates
is \be P_- \left( P_+ + 2 \sum_i\beta_i L_i \right) + \sum_i
\left( \dot{\rho}_i^2 + \beta_i^2 P_-^2\rho_i^2 +
\frac{L_i^2}{\rho_i^2} \right) = - (\ep + P^2). \label{eqn:L} \ee
 
The radial equations are \be \ddot{\rho}_i = \frac{1}{\rho_i}
\left( \frac{L_i}{\rho_i} - \beta_i P_-\rho_i \right) \left(
\frac{L_i}{\rho_i} + \beta_i P_- \rho_i \right), \ee
 
\noindent and, since the space is homogeneous, with no loss of
generality we need consider only those geodesics which pass
through the origin of our coordinate system.  Just as in the
GLU~\cite{boyda:02}, it is clear from the form of the effective
potential in (\ref{eqn:L}) that this requires $L_i=0$.  We
therefore obtain the following simple solutions for the radii: \be
\rho_i(\lambda) = \rho_i^{max} \sin ( \beta_i P_- \lambda),
\label{eqn:radial} \ee
 
\noindent where we have chosen those geodesics which pass through
the origin at $\lambda = 0$.  Solving (\ref{relations}) for the
remaining coordinates then gives \ba t(\lambda) &=& t^{(0)} +
\frac{E}{2} \lambda - \frac{1}{2} \sum_i \beta_i (\rho_i^{max})^2
\left( \beta_i P_- \lambda - \frac{1}{2} \sin(2 \beta_i P_-
\lambda)
\right), \nonumber \\
z(\lambda) &=& z^{(0)} + \frac{P_z}{2} \lambda + \frac{1}{2}
\sum_i \beta_i (\rho_i^{max})^2 \left( \beta_i P_- \lambda -
\frac{1}{2} \sin(2\beta_i P_- \lambda)
\right), \label{eqn:coords} \\
\p_i (\lambda) &=& \p_i^{(0)} + \beta_i P_- \lambda, \nonumber \ea
 
The behaviour of the geodesics is similar to those in the
GLU~\cite{boyda:02}.  They spiral out from the origin, approach
some maximum radius, $\rho_i^{max}$, when $\la = \pi/(2\beta_i
P_-)$, and re--focus back at the origin in another $\pi/(2\beta_i
P_-)$ of affine distance.  One can also see that $t$ is not a good
affine parameter for these geodesics.  The important difference,
however, is the value of the maximum radii for timelike and null
geodesics.  We now show that these geodesics can probe the region
beyond the critical radii where the CTCs develop.
 
The constraint (\ref{eqn:L}) with $L_i=0$ can be used to put
limits on $\rho_i^{max}$. Consider for simplicity the geodesics
which remain at the origin for all the planes but one\,\footnote{In
the spherically symmetric case, for which $\beta_i = \beta ~
\forall i$, other geodesics are related to these by symmetry
generators.}, that is $\rho_i^{max} =0 \,\,\,\mbox{for} \,\,\,i
\neq 1$. Then, denoting $\beta_1 = \beta$ and $\rho_1^{max} =
\rho_{max}$, we have \be \rho_{max} = \frac{1}{\beta}
\sqrt{\frac{-(\ep + P^2 + P_+ P_- )}{P_-^2}}. \ee
 
\noindent It is clear that $\rho_{max}$ is largest (for causal
propagation) when the geodesics are null, for which $\ep=0$, and
have vanishing momenta in the flat directions.  For those
geodesics one obtains \be \rho_{max} = \frac{\alpha}{\beta} =
\frac{1}{\beta} \sqrt{ \frac{E-P_z}{E+P_z} }, \ee where we have
defined $\alpha = \sqrt{-P_+/P_-}$.  Note that $|P_z| \le E$ gives
$0 \le \alpha < \infty$.
 
The behaviour of the geodesics in the $\{t,\rho,\p\}$ directions
in shown in figure 1 where in this, and all other, plots we have
used the dimensionless variables $\beta P_- \la, \,\beta t,
\,\beta z$ and $\beta \rho$. The Kaluza--Klein zero modes, with $P_z=0
~(\alpha=1)$,
reach a maximum radius of $\beta^{-1}$, just as in the discussion
of the GLU~\cite{boyda:02}.  However, the non--zero modes, with
$P_z < 0$, have $\rho_{ max} > \beta^{-1}$.    We therefore
conclude that on--shell physical trajectories can probe those
regions of spacetime in which CTCs appear.  Indeed, for $\alpha$
large enough, the trajectories actually run backwards in time for
some range of affine parameter.

\[
\vbox{
         \beginlabels\refpos 0 0 {}
                     \put 160 -10 {t^{(0)}+\pi/2}
                     \put 128 -116 {\rho}
                     \put 162 -141 {\p}
                     \put 192 -222 {t^{(0)}}
         \endlabels
         \epsfxsize=.4\hsize
         \centerline{\epsfbox{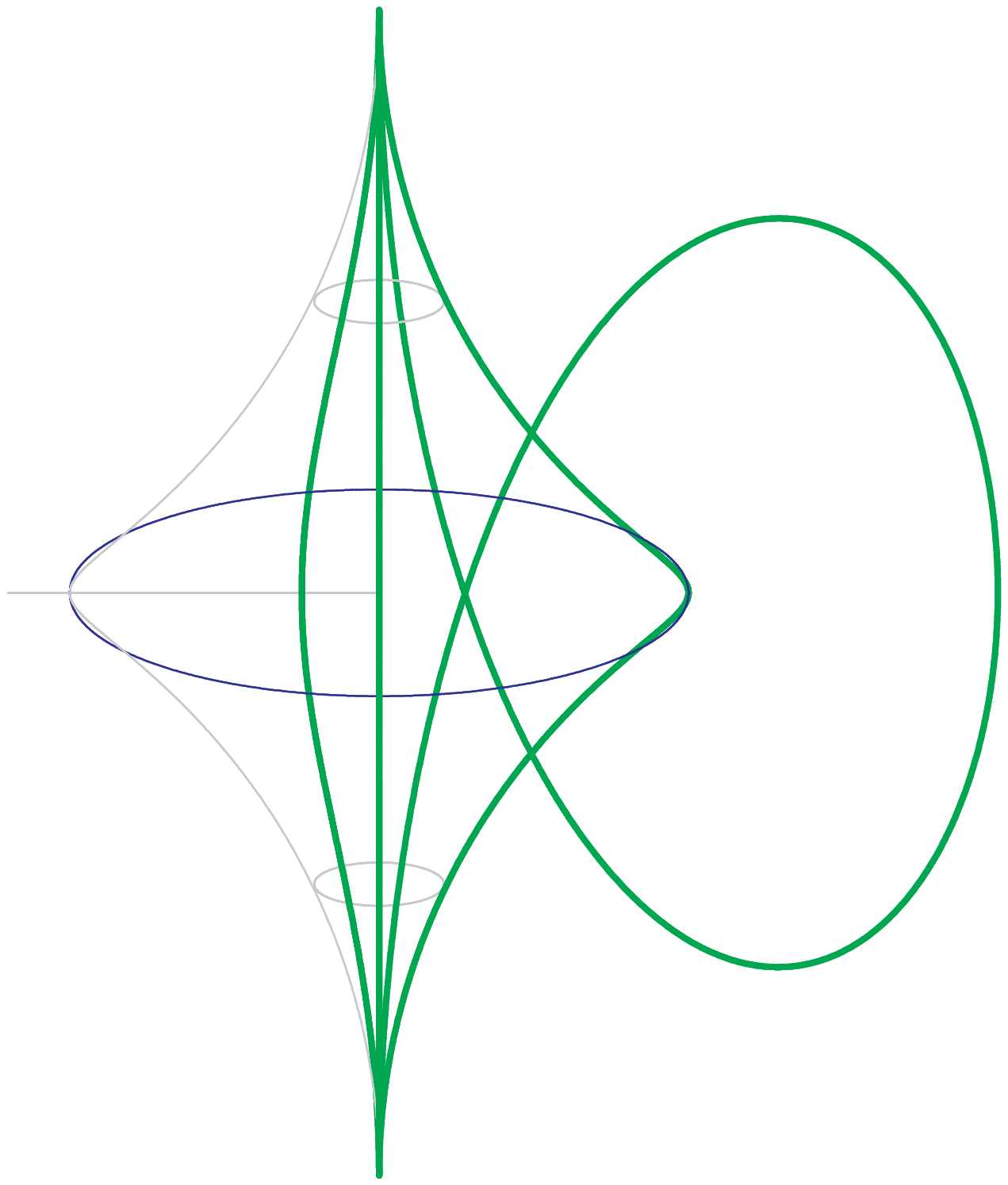}}
         {\footnotesize Figure 1.  Null geodesics for different values of
         $\alpha$ are plotted as thick green lines.  Starting at some
         $t=t_0$, they re--focus at $t=t^{(0)} + \pi/2$, reaching a maximal
         radius when $\la = \pi/(2\beta P_-)$.  For $\alpha > 1$,
         geodesics can probe radii at which CTCs appear, represented
         by the blue circle of radius $\rho = \beta^{-1}$.}
            }
\]

Of course, the geodesics also move in the (compact) $z$ direction,
and we exhibit this motion in figure 2, where $z$ runs around the
circle, and $\rho$ along the axes of the cylinder.  Each point on
the surface of this cylinder is a circle of radius $\rho$.  Note that
the period of $\beta z$ is $2\pi \beta R$, $\beta R$ being a
dimensionless parameter characterizing the geometry.  For
the purposes of this, and all other, plots, we have (arbitrarily)
chosen $\beta R= 1$.
 
The maximal radii for causal propagation depend on the
Kaluza--Klein momentum $P_z$. This is represented in the reduced
G\"{o}del metric as an electric charge. The null geodesics in the
CPW geometry would appear timelike in the reduced GLU geometry,
representing trajectories of massive BPS particles. The
interaction with the background magnetic field will be responsible
for the different behaviour in that language.

\[
\vbox{
         \beginlabels\refpos 0 0 {}
                     \put 142 -13 {\rho=\infty}
                     \put 142 -135 {\rho=\beta^{-1}}
                     \put 142 -168 {\rho=0}
                     \put 260 -180 {z^{(0)}}
                     \put 273 -152 {z^{(0)}+\pi/2}
         \endlabels
         \epsfxsize=.2\hsize
         \centerline{\epsfbox{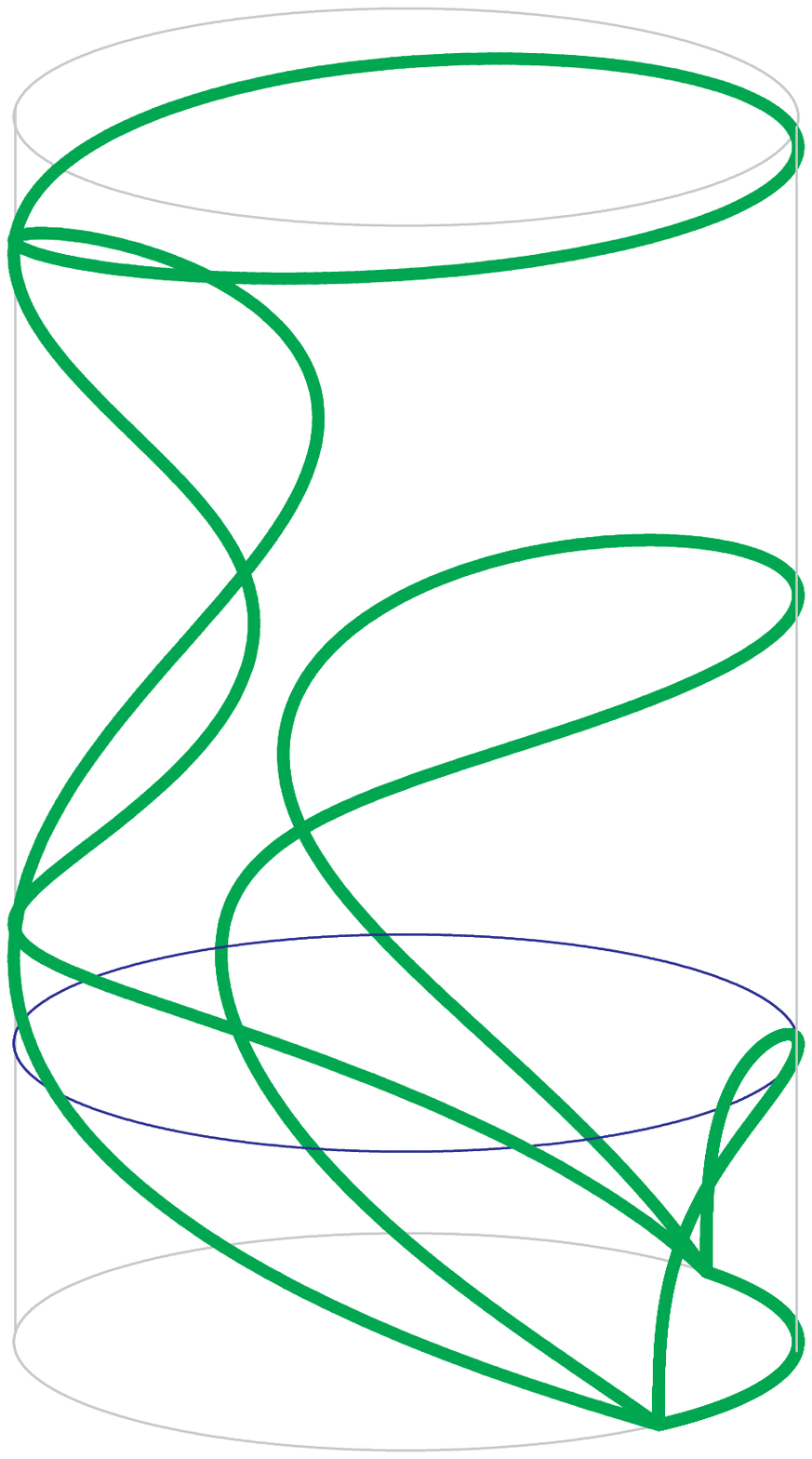}}
         {\footnotesize Figure 2.  Geodesics for different
         values of $\alpha$ are plotted as thick green lines.  Starting
at
         $\rho=0$ and $z=z^{(0)}$, they re--focus at $\rho=0,
         z=z^{(0)} + \pi/2$, reaching a maximal radius for $\la =
\pi/(2\beta
         P_-)$.  For $\alpha > 1$, geodesics can probe radii at which
         CTCs appear, represented by the blue circle at $\rho =
         \beta^{-1}$.}
            }
\]
 
Having found the general geodesic in the CPW, we can now prove
that there are no closed timelike or null geodesics in this
background. To see this, consider the identification of
coordinates in the CPW: \be (u , v, \p_i) \equiv (u + 2 \pi n R, v
+ 2 \pi n R, \phi_i + 2 \pi m_i) \ee
 
\noindent Due to the homogeneity of the geometry it is sufficient
to consider a timelike or null geodesic starting from
$(u,v,\rho,\phi_i) = (0,0,0,0)$, as above.  The question is then
whether this geodesic can pass through the point $(2 \pi n R,2 \pi
n R, 0,2 \pi m_i )$.  Given the equations for geodesics,
(\ref{eqn:radial}) and (\ref{eqn:coords}), this would require \ba
P_- \lambda &=& 2 \pi n R \nonumber \\
P_- \beta \lambda &=& 2 \pi m_i \\
-\frac{(\ep + P^2) \la}{P_-} &=& 2 \pi n R = P_- \lambda \nonumber
\ea
 
\noindent Note that the second equation requires $\beta$ to be a
rational multiple of $1/R$ --- this is not required for CTCs but
is necessary for closed geodesics.  The third equation gives $-\ep
= P_-^2 + P^2$, which can never be satisfied for timelike
geodesics, and can only be satisfied for null geodesics when $P_-
= P^2 = 0$, which does not lead to a closed geodesic.  Thus, the
only closed geodesics are spacelike as promised.


\sect{Preferred Holographic Screens}
 
To construct the preferred holographic screens, we first discuss the
construction of lightsheets and their bounding 
surfaces. As the
geometry can be made only mildly curved, it is a perfect candidate
for application of the ideas of \cite{bousso:9905,bousso:9906}.
 
Since the space is homogeneous, we can concentrate on screens
constructed from geodesics which emanate from the point we call our
origin. The resulting screens
will be observer--dependent.  The holographic
screen~\cite{bousso:9905,bousso:9906} is defined by taking a
congruence of future-- or past--directed null geodesics at each
point on the observer's worldline, and following them along the
direction in which the expansion $\theta$ of that congruence is
positive.  To construct a \emph{preferred} screen, we terminate
the geodesics at the points for which $\theta$ vanishes.
 
We first discuss the construction for a fixed point in time along the observer's
worldline. This
results in a light--sheet for the observer. The covariant entropy
bound~\cite{bousso:9905} states that the entropy on this light
sheet is bounded by the area of the orthogonal spatial
surface\,\footnote{We refer to a codimension one object as a ``screen'',
and a codimension two object as a ``surface''.} for which $\th = 0$.
The holographic screen is then simply the union of these surfaces
for each point in time along the observer's worldline.
 
To simplify matters, we set $\beta_i = \beta ~ \forall i$ and
introduce an overall radial coordinate $r^2 = \sum_i
\rho_i^2$ and some angles $\th_m, m=1, \ldots , N-1$ to replace
the $\rho_i$.  We write $\rho_i = r \hat{\rho}_i$ where the
direction cosines satisfy $\sum_i \hat{\rho}_i^2 = 1$.  In
this coordinate system, the geodesics discussed above, namely the
future--directed null geodesics which pass through the origin and
which vary only in one plane, have $\th_m = {\rm constant}$ and
\be r (\la) = \frac{1}{\beta} \sqrt{\al^2 - \frac{P^2}{P_-^2}}
\sin(\beta P_- \la). \ee
 
The above construction entails sending out null geodesics from the
origin at each moment in time and in \emph{all} directions.  For
fixed $t$, the surfaces at constant $\la$ will thus be codimension
two, as required.  We terminate the geodesics when $\theta = 0$.
Let us now drop the flat directions, taking $d$ to be the
dimension of the resulting GLU space\,\footnote{We can easily
construct a screen relevant to an observer delocalized in the flat
directions, by taking $P^2=0$ and then smearing over these
directions. (What we mean by this process of smearing will be
discussed in the following section.)  Accounting for non--zero
$P^2$ complicates matters considerably, as in the analysis
of~\cite{boyda:02} for the GLU.}.  Then the surface at $\theta=0$
is parameterized by the $d-2$ directions in which the geodesics can
leave the origin. By construction, a set of coordinates for this
surface is given by $\{\al,\th_m,\p^{(0)}_i \}$.  Coordinates on
the screen are then $\{t^{(0)},\al,\th_m,\p^{(0)}_i \}$.  One can,
of course, compute $\theta$ in any coordinate system, but it is
easiest to use these coordinates, which are adapted to the null
congruence.
 
We are free to set $P_- = 1$ by rescaling $\la$\,\footnote{One can
show that the location of the surface $\theta = 0$ is independent of
the choice of $\la$ under such rescalings.}, in which case the
spacetime metric is
\[
\d s^2 = -(1+\alpha^2) \d t^{(0)} \d\la -(\d t^{(0)})^2 - 2
\frac{\al}{\beta} \sin(\beta \la) \left(
\cos (\beta \la) \d \al + \al \sin(\beta \la) \sum_{i=1}^N
\hat{\rho}_i^2 \d \p_i^{(0)} \right) \d t^{(0)}
\]
\be
+ \frac{\sin^2 (\beta
\la)}{\beta^2} \left( \d \al^2 + \al^2 \d \Om_{d-3}^2 \right).
\label{eqn:geo_metric}
\ee
 
\noindent Note that the tangent
\be
\xi = \frac{\del}{\del \la},
\ee
 
\noindent to the geodesics is manifestly null, and that the
coordinates on the surface $\theta = 0$, to be exhibited shortly,
are manifestly orthogonal to this tangent.
 
The expansion of the congruence is particularly simple:
\be
\th = D \cdot \xi = (d-2) \beta \cot(\beta \la),
\label{eqn:theta}
\ee
 
\noindent so that the surface for which $\th=0$ is thus defined by
\be \la = \frac{1}{\beta} \frac{\pi}{2}.
\label{eqn:lambda} \ee
 
\noindent In terms of the original coordinates a point on the
surface is given by \be
z = z^{(0)} + \frac{1}{\beta} \frac{\pi}{4}, \qquad
r = \frac{\al}{\beta}, \qquad
\p_i = \p_i^{(0)} + \frac{\pi}{2}, \qquad
\th_m = {\rm constant}.
\ee

The induced metric on the holographic screen is given by
substituting the condition
(\ref{eqn:lambda}) on $\la$ into the metric (\ref{eqn:geo_metric}),
giving
\be
\left. \d s^2 \right|_{screen} = -(\d t^{(0)})^2 - 2 \alpha^2
\beta \sum_{i=1}^N \hat{\rho}_i^2 \d \p_i^{(0)} \d t^{(0)} + \d \al^2
+ \al^2 \d \Om_{d-3}^2,
\ee
 
\noindent where we have rescaled $\al$ by $\beta$.  The screen has
Lorentzian
signature, with each spacelike slice at constant $t^{(0)}$ being
flat $\mathbb{E}^{d-2}$, $\al$ acting as the radial coordinate in
this space.  It seems to exhibit rotation, but defining the
coordinates \be \hat{\p}^{(0)}_i = \p^{(0)}_i - \beta t^{(0)}, \ee
 
\noindent gives \be \left. \d s^2 \right|_{screen} = - (1 + \al^2
\beta^2 ) (\d t^{(0)})^2 + \d\al^2 + \al^2 \d \Om_{d-3}^2, \ee
 
\noindent where now the $(d-3)$--sphere is parameterized by
$\{\hat{\p}^{(0)}_i,\th_m\}$.  For small $\al$, the induced metric
is that of flat Minkowski space.  Note that the area of the
surface bounding the lightsheet, at constant $t^{(0)}$, is infinite;
it is just flat space.  The covariant entropy
bound~\cite{bousso:9905,bousso:9906} thus implies that the entropy
enclosed on the lightsheet can be infinite, as in AdS and
Minkowski space, but unlike deSitter space.

To picture the holographic screen in spacetime, it is useful to first
consider the case of the non--compact plane
wave (the compact case can then be analysed by identifying points
the $z$ direction).  Indeed, given the difficulties in understanding holography
in the plane wave~\cite{pp1}--\cite{pp6}, this is a useful exercise in
itself.  All the previous analysis can be applied directly to the
non--compact plane wave.  In particular, the expressions
(\ref{eqn:radial}), (\ref{eqn:coords}) and (\ref{eqn:theta}) for the
null geodesic congruence and its expansion remain valid.  Of course,
there are no issues concerning chronology protection in this case
since the non--compact plane wave does not contain CTCs.

Figure 3 shows the preferred holographic screen in the non--compact
geometry.  Since the screen extends over
the entire range of the angular coordinates $\{\p_i,\th_m\}$ and
time $t$, we are only concerned with its position in the $r$
and $z$ directions.  It is represented by the
red line, parameterized by the coordinate $\alpha$. The
thick green lines are the null geodesics which terminate at the
screen, at a value of $\la$ given by (\ref{eqn:lambda}).  At fixed
$t^{(0)}$, these
geodesics form one of the lightsheets for the screen, and it is easy
to see that taking the collection of these lightsheets for all
$t^{(0)}$ fills the entire region to the left of the
screen; it is just the region $z < \pi/4$.  We could alternatively
consider the lightsheets which terminate at the caustic
$z=z^{(0)} + \pi/2$.  These would cover the region $z > \pi/4$.

When discussing the non--compact plane wave, one might prefer to construct
the holographic screen in light--cone coordinates, using $u^{(0)} =
\left( z^{(0)} + t^{(0)} \right)/2$ as the time
coordinate on the screen.  In these coordinates, the condition
(\ref{eqn:lambda}) for vanishing $\theta$ still holds.  The induced
metric on the screen is
\be
\d s^2 = - \al^2 \beta^2 (\d u^{(0)})^2 + \d\al^2 + \al^2 \d
\Om_{d-3}^2,
\ee

\noindent and a point on the surface of the screen is given by \be
v = v^{(0)}, \qquad r = \frac{\al}{\beta}, \qquad
\p_i = \p_i^{(0)} + \frac{\pi}{2}, \qquad
\th_m = {\rm constant}.
\ee

\noindent If we replace $z$ with $v$ in figure 3, and move the screen
to $v^{(0)}$, then a similar picture would still be valid.  

It would be interesting to find a relation between this preferred
screen and the boundary of the plane wave, which was found to be a null
line in~\cite{pp4,pp5}.

The case of the \textit{compact} plane wave is more complex.  To
discuss this (for the case $\beta =1/R$), we simply identify the
lines $z=z^{(0)} + 2\pi n$.  Then circles on the
cylinder at constant $r > \beta^{-1}$ will be the projection into
the $\{r,z\}$ plane of one of the CTCs discussed in section 2.

In the GLU, as in other interesting geometries, the surface $\theta = 0$
encloses a region of space, so that the spacelike projection
of~\cite{bousso:9905,bousso:9906} can be applied.  In this
sense, the concept of holographic protection of chronology is
quite appealing: spatial sections of the screen (surfaces) should
encode all information, or degrees of freedom, within a volume of
space.  An holographic theory on the screen would describe physics
in a region of spacetime which does not contain CTCs.
 
In our case, it is easy to see that any point in spacetime is
``enclosed'' within the holographic screen.  The collection of
lightsheets shown in
figure 3 for the CPW covers the whole of spacetime.  Thus, the region
``cut out'' by the screen is not free of causal ambiguity.
Holographic protection of chronology in this case does not appear to
be in operation.

\[
\vbox{
         \beginlabels\refpos 0 0 {}
                     \put 110 -10 {r=\infty}
                     \put 110 -200 {r=\beta^{-1}}
                     \put 110 -230 {r=0}
                     \put 275 -243 {z^{(0)}}
                     \put 300 -246 {z^{(0)}+\pi/4}
                     \put 180 -246 {z^{(0)}-2\pi}
         \endlabels
         \epsfxsize=.35\hsize
         \centerline{\epsfbox{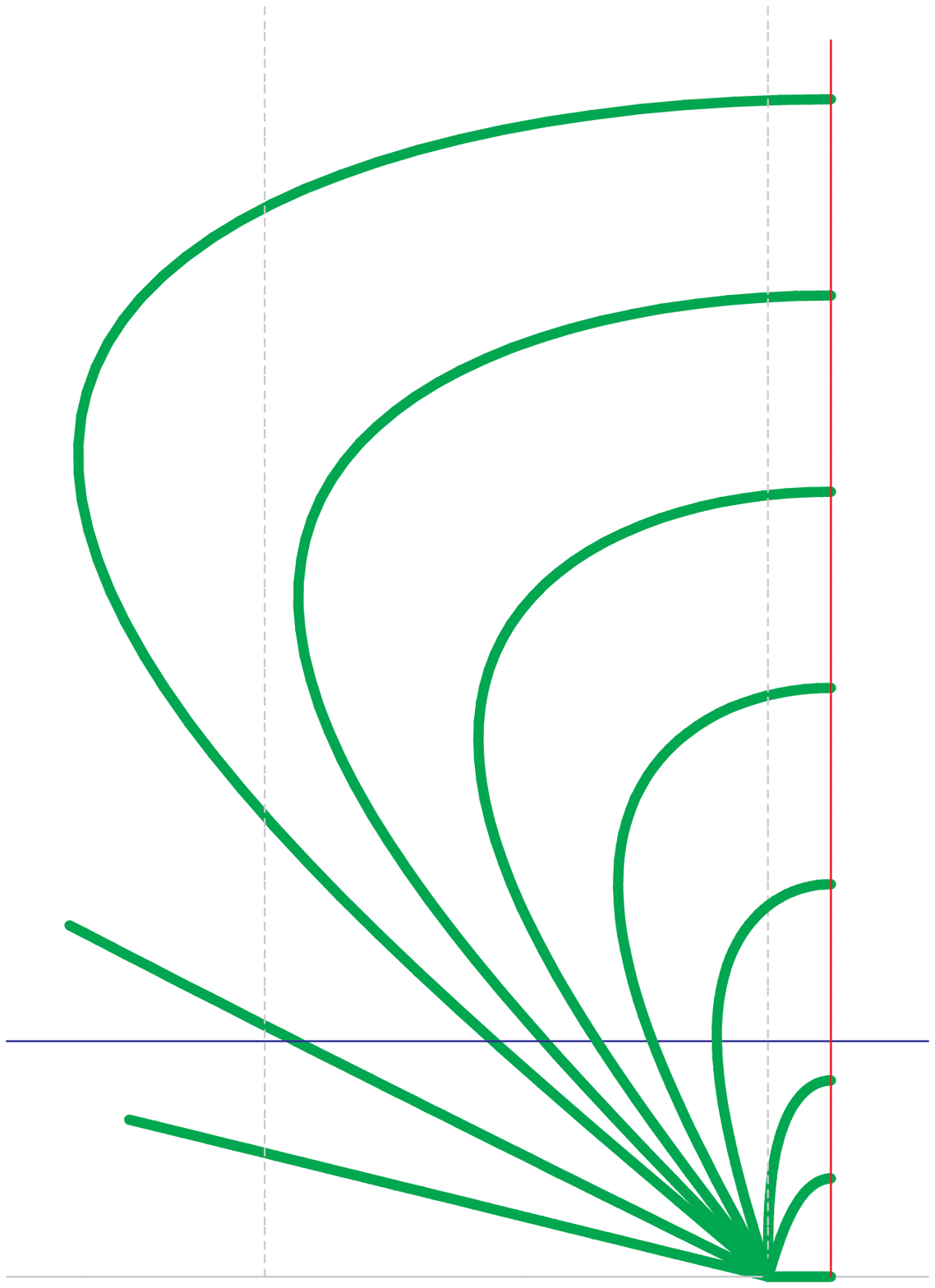}}
         \vspace{15pt}
         {\footnotesize Figure 3.  The holographic screen is
         shown in red, $\alpha$ being the coordinate along this line
         of constant $z = z^{(0)} + \pi/4$,
         and each point being a sphere of
         radius $r$.  Geodesics for different values of $\alpha$ are
         again plotted as thick green lines, starting from the origin
         at $z=z^{(0)}$.  For constant $r$, the screen
         is just a point in the $z$ direction.  To discuss the compact
         plane wave (with $\beta = 1/R$), we simply identify
         $z=z^{(0)}$ with the dotted lines $z=z^{(0)} \pm 2n\pi$.  In
         this case, the blue line represents the projection of a CTC
         into the $\{r,z\}$ plane.}
            }
\]


\sect{Relation to the G\"{o}del--Like Universe}
 
When the geometry contains a Killing vector along a possibly compact
direction, as in the GLU with
flat transverse directions~\cite{boyda:02}, there are at least two
types of lightsheet and screen one can discuss.  The first is that
which we have constructed above, and is relevant to an observer
localized in all directions.  It is not translationally invariant
along the compact direction, so it is difficult to see how it is
connected to the lower--dimensional screen in the GLU.
 
However, as in~\cite{boyda:02}, one may also contemplate
``smearing'' in the compact directions.  These latter screens are
constructed by taking $\al = 1$ or $P_z = 0$, \emph{i.e.}, for the
Kaluza--Klein zero modes.  Of course, $\al$ is then no longer any
use as a coordinate along the screen, but since \be
\left(\frac{\del}{\del z^{(0)}}, \xi \right) =
\left(\frac{\del}{\del z}, \xi \right)  = \frac{1}{2} (1-\al^2),
\ee
 
\noindent the vector $\del/\del z_0$ is orthogonal to the tangent
$\xi$ when $\al = 1$.   For these light rays, $z_0$ replaces $\al$
as a coordinate on the screen.  Working with these coordinates, one
finds that the expansion is given by
\be
\theta =\frac{\left((d-3)-(d-2)\beta^2 r^2
\right)}{r\sqrt{1-\beta^2r^2}},
\ee

\noindent so that
\be
r = \sqrt{\frac{d-3}{d-2}} ~\frac{1}{\beta} < \frac{1}{\beta} \qquad
\Leftrightarrow \qquad
\sin(\beta \la) = \sqrt{\frac{d-3}{d-2}},
\ee

\noindent is the surface $\theta = 0$.  The smeared screen thus
provides a manifest holographic protection of chronology.  Note
also that, unlike the unsmeared case, all null geodesics intersect
the smeared screen precisely twice in each cycle (as $\lambda$
runs from $0$ to $\pi/(2\beta P_-)$).

It is only this smeared screen which is relevant to the GLU. It is
translationally invariant along $z$, so can be dimensionally
reduced in the usual manner.  Indeed, it is easy to see that the
induced metric on the smeared screen is \ba \left. \d s^2
\right|_{screen} &=& - \left( \d t^{(0)} +
\frac{1}{\beta}\frac{d-3}{d-2} \sum_{i=1}^N \hat{\rho}^2_i \d
\p^{(0)}_i \right)^2 + \frac{1}{\beta^2} \frac{d-3}{d-2} \d
\Om^2_{d-3} \nonumber
\\
&& \qquad \qquad \qquad \qquad \qquad + \left( \d z^{(0)} -
\frac{1}{\beta} \frac{d-3}{d-2} \sum_{i=1}^N \hat{\rho}^2_i \d
\p^{(0)}_i \right)^2, \label{eqn:smeared} \ea
 
\noindent which one can dimensionally reduce along $z^{(0)}$ to
give the induced metric on the screen in the GLU~\cite{boyda:02},
or T--dualize to give the induced metric on the smeared screen in
the GLU times a circle.
 
A similar structure exists for  a more general Kaluza--Klein
dimensional reduction.  Denoting the $(D-1)$--dimensional
coordinates by $x^a$, the $z$ equation of motion for a particle
moving in the $D$--dimensional metric \be \d s^2_D = e^{2\al \p}
\d s^2_{D-1} + e^{2 \beta \p} ( \d z + A )^2, \ee
 
\noindent is $\dot{z} = e^{-2\beta \p} P_z - A_a \dot{x}^a$.  The
vector tangent to a null geodesic (in the higher dimensional
sense)
 will have the form \be \xi = \dot{x}^a \frac{\del}{\del
x^a} + \dot{z} \frac{\del}{\del z}, \ee
 
\noindent so that \be \left( \frac{\del}{\del z^{(0)}}, \xi
\right) = \left(\frac{\del}{\del z}, \xi \right) = P_z. \ee
 
\noindent The localized screen, the position of which varies in the
$z$ direction, is parameterized by $P_z$. There also
exists a  smeared screen, however, which is parameterized by $z^{(0)}$,
and
for which $P_z=0$.
 
The induced metric on the smeared screen is a generalization of
(\ref{eqn:smeared}): \be \left. \d s^2_D \right|_{screen_D} =
e^{2\al \p} \left. \d s^2_{D-1} \right|_{screen_{D-1}} + e^{2
\beta \p} ( \d z^{(0)} + A )^2, \ee
 
\noindent where the fields $\p$ and $A$ should be evaluated on the
$(D-1)$--dimensional screen.  We see that the areas of the screens
in the corresponding Planck units behaves simply under dimensional
reduction, if the higher dimensional screen is smeared.  Similar
comments apply to geometries related by T-duality in which, at least
for the case at hand, one can think of the Kaluza--Klein vector as the
$z$ components of the $B$--field.  We note that
the lower dimensional Planck scale is determined at the screen's
location, in case the dilaton varies.
 
The relation between the localized screens in $D$ and $D-1$
dimensions (or equivalently between the localized screen and its
smeared version) is unclear.  In particular, the entropy bound
derived from the lower dimensional surfaces (\emph{i.e.} sections of the
corresponding screen) may be stronger or weaker than the covariant
entropy bound of the full geometry.  One may be tempted to regard
the lower dimensional version as more  coarse grained and the
associated entropy as excluding the massive Kaluza--Klein modes; it would be
useful to make this more precise.


\sect{Conclusions}
 
Perhaps the most interesting question concerning the spaces which
we have been discussing is whether string theory can, in some
sense, ``cope'' with CTCs.  Certainly, the very existence of
highly supersymmetric string backgrounds with CTCs suggests that
such geometries should not \emph{a priori} be excluded from
acceptable string vacua without a concrete physical reason. In the
present context, due to special features of these backgrounds, it
is still unclear as to whether quantum field theory, or string
theory,  can be consistently defined\,\footnote{For various
efforts in the four--dimensional context,
see~\cite{leahy:82,novello:92}.}.
 
Neither the GLU nor the CPW are globally hyperbolic, the classical
Cauchy problem being ill--posed in both spaces\,\footnote{In the
GLU, this is intimately related to the existence of CTCs, but the
lack of a Cauchy surface in the case of the CPW runs deeper: even
the uncompactified plane wave, with no CTCs, is not globally
hyperbolic~\cite{penrose:65}.  Defining quantum field theory on
this space is thus a somewhat subtle issue.  The surface $u={\rm
constant}$ can be used as a substitute Cauchy surface as in,
\emph{e.g.},~\cite{brecher:02}.  This fails to capture only those
geodesics which travel parallel to the wave~\cite{gibbons:75}.
Alternatively, as we are arguing here, one can impose a cutoff on
the transverse coordinates, leaving a patch of spacetime for which
the classical Cauchy problem is well--defined~\cite{marolf:02}.}.
As in AdS spaces, one possible approach to define dynamics on such
a spacetime is to supplement the initial conditions with some
boundary data.  The crucial issue is whether there exists a choice
of such boundary conditions which renders the resulting
propagation consistent.
 
Unlike the AdS case, here the surface on which to define boundary
data is most naturally the Cauchy horizon.  Such a
prescription in other instances leads to instabilities,
\emph{e.g.}, for the Cauchy horizon in the interior of generic
black holes~\cite{inner,Ross,charged}.  Here no such instabilities
manifest themselves. This is related primarily to two important
facts: the spectrum of matter fluctuations is discrete  (a fact
inherited from the plane wave), and the CTCs are not geodesics.
 
One can, for example, expect the energy--momentum tensor of matter
fields to diverge at the Cauchy horizon. This is not expected to be the
case
here, as the spectrum of modes is discrete. Related to that, a
possible classical super--radiance phenomena would be a signal of
instability. This requires separation of modes into incoming and
outgoing, which is unnatural in a plane wave geometry.
 
In case of geodesic CTCs, such as in flat space with compact
time, one can easily construct problematic amplitudes using the
fact that CTCs dominate some quantum amplitudes in a
semi--classical approximation. In our case, it is not clear that
the presence of off--shell trajectories which are CTCs generates
any problem for quantum mechanics in this space.
 
Recently, practical methods have been developed to deal with the
CTCs present in certain orbifolds~\cite{biswas:03}.  Although
interesting, it is not clear as to what extent such methods would
be applicable here: much use is made of the covering space of the
orbifold, a notion which does not generalize to the case at hand.
 
Alternatively~\cite{hawking:95}, at least in the case of the
GLU~\cite{radu:98,radu:01}, one can analytically continue to the
Euclidean regime and use the techniques of Euclidean field theory.
Results thereby obtained can be analytically continued back to the
Lorentzian regime to obtain information about quantum fields in
the original space. For additional discussion of QFT in
non-globally hyperbolic spaces see,
\emph{e.g.},~\cite{kay92,wald1,wald2}.
 
Using holography in these backgrounds is appealing,
especially due to the need to carve out a finite region of spacetime
in order to define the dynamics uniquely.  However, at
least in one example, we have found that the concept of
``holographic protection'' of chronology appears to be inadequate.  It
would be interesting to examine further, and in more generality, the connections
between the concepts of holography and chronology protection.  It
would also be of interest to examine more quantatively how the
processes of dimensional reduction and/or T--duality affect the
covariant entropy bounds of Bousso~\cite{bousso:9905,bousso:9906}.
 
One may be tempted to use relations to uncompactified plane waves, or to
AdS~\cite{squashed}, to find the holographic theory living on the
screens directly.  Such attempts are not straightforward, as is
demonstrated by our lack of understanding of plane wave
holography~\cite{pp1}--\cite{pp6}.


\vspace{1cm} \noindent
{\bf Acknowledgments}\\
We are grateful to Ed Boyda for correspondence, and would like to
thank Rob Myers, Kristin Schleich, Gordon
Semenoff, Bill Unruh, Uday Varadarajan and Don Witt for useful
conversations.  DB also thanks Paul Saffin for early collaboration
and Joan Sim\'{o}n for useful correspondence concerning CTCs in
CPWs.  DCP thanks Dumitru Astefanesei, Kazuo Hosomichi, Amanda Peet,
Simon Ross and Konstantin
Savvidis, and is grateful to the Perimeter Institute for
hospitality during the completion of this paper.  This work is
supported in part by NSERC, and DCP is funded by Ontario PREA.



\end{document}